\documentclass[twocolumn]{aastex61}

\begin{document}

\title{\emph{SWIFT} Detection of a 65-Day X-ray Period from the Ultraluminous Pulsar NGC 7793 P13}

\newcommand{\swift}{\emph{Swift}}
\newcommand{\ngc}{{NGC 7793 P13}}
\newcommand{\LD}[1]{{\color{red}\bf #1}}
\newcommand{\BD}[1]{{\color{black} #1}}

\correspondingauthor{Chin-Ping Hu, K. L. Li}
\email{cphu@hku.hk, liliray@pa.msu.edu}

\author{Chin-Ping Hu}
\affiliation{Department of Physics, The University of Hong Kong, Pokfulam Road, Hong Kong}
\author{K. L. Li}
\affiliation{Department of Physics and Astronomy, Michigan State University, East Lansing, MI 48824, USA}

\author{Albert K. H. Kong}
\affiliation{Institute of Astronomy and Department of Physics, National Tsing Hua University, Taiwan}
\affiliation{Astrophysics, Department of Physics, University of Oxford, Keble Road, Oxford OX1 3RH, U.K.}

\author{C.-Y. Ng}
\affiliation{Department of Physics, The University of Hong Kong, Pokfulam Road, Hong Kong}

\author{Lupin Chun-Che Lin}
\affiliation{Institute of Astronomy and Astrophysics, Academia Sinica, Taiwan}

\begin{abstract}
NGC 7793 P13 is an ultraluminous X-ray source harboring an accreting pulsar. We report on the detection of a $\sim$65\,d period X-ray modulation with \swift\ observations in this system. The modulation period found in the X-ray band is $P=65.05\pm0.10$ d and the profile is asymmetric with a fast rise and a slower decay. On the other hand, the \emph{u}-band light curve collected by \swift\ UVOT confirmed an optical modulation with a period of $P=64.24\pm0.13$ d. \BD{We explored the phase evolution of the X-ray and optical periodicities and propose two solutions}. A superorbital modulation with a period of $\sim$2,700--4,700\,d probably caused by the precession of a warped accretion disk is necessary \BD{to interpret the phase drift of the optical data}. \BD{We further discuss the implication if this $\sim$65\,d periodicity is caused by the superorbital modulation. Estimated from the relationship between the spin-orbital and orbital-superorbital periods of known disk-fed high-mass X-ray binaries, the orbital period of P13 is roughly estimated as 3--7\,d. In this case, an unknown mechanism with a much longer time scale is needed to interpret the phase drift.} Further studies on the stability of these two periodicities with a long-term monitoring could help us to probe their physical origins.
\end{abstract}


\section{Introduction}
Ultraluminous X-ray sources (ULXs) are non-nuclear and point-like sources with X-ray isotropic luminosities higher than the Eddington limit of a $\sim$10\,$M_{\odot}$ black hole ($\gtrsim10^{39}$\,erg\,s$^{-1}$) in nearby galaxies. The apparent luminosity of an accreting black hole can be inferred from three parameters: the mass accretion rate $\dot{m}$, black hole mass $M_{\rm{BH}}$, and the beaming factor $b$ \citep{ShakuraS1973, PoutanenLF2007}. Therefore, ULXs were thought to harbor stellar mass black holes with super-Eddington accretion rate and mild beaming, or intermediate mass black holes ($M\sim10^2$--$10^4\,M_{\odot}$) in a sub-Eddington accretion regime \citep[see e.g.,][]{FengS2011}. However, our understanding of ULXs was challenged by the discovery of an accreting neutron star in ULX M82~X-2 \citep{BachettiHW2014}. Recently, two ultraluminous pulsars, \ngc\ and NGC 5907 ULX-1, were identified \citep{IsraelPE2016,FuerstWH2016,IsraelBS2016}. These discoveries imply that a non-negligible number of ULXs may host neutron stars. 

\begin{figure*}
\plotone{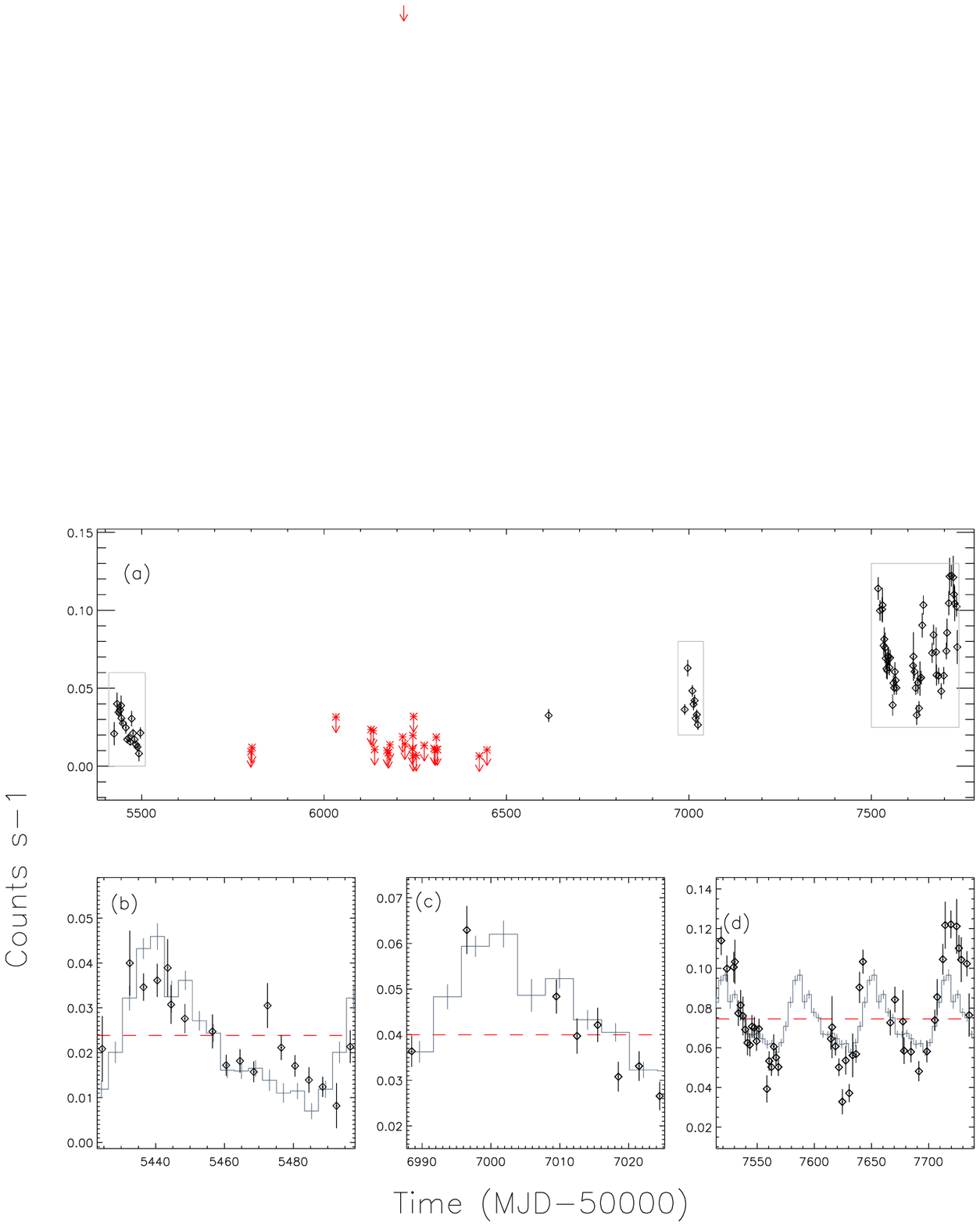}
\caption{(a) \swift\ XRT 0.3--10\,keV light curve of \ngc\ binned to 1-day. The red arrows mark the $3\sigma$ upper limits. Gray boxes indicate three segments of clustering data points. (b)--(d): Zoom-in view of each segment. Black diamonds, red dashed lines, and gray histograms are the data points, mean values of corresponding segments, and the X-ray light curve folded with 65.05\,d, respectively. \label{lc_xrt}}
\end{figure*}

\ngc\ (hereafter P13) was identified by \emph{ROSAT} as the brightest X-ray point source in NGC 7793 at an X-ray luminosity of $(1.4-1.8)\times10^{39}$\,erg\,s$^{-1}$ \citep{ReadP1999} with a distance of $D=3.6$--$3.9$\,Mpc \citep{KarachentsevGS2003, Radburn-SmithJS2011, TullyCS2016}. Further deep \emph{Chandra} observation revealed two X-ray sources within the \emph{ROSAT} PSF. The much brighter one, CXOU J235750.9$-$323726, is associated with the ULX P13 \citep{Pannuti2011}. The optical counterpart was classified as a B9Ia star with a \emph{V}-band magnitude of $\sim$20.5 \citep{MotchPG2011}. Optical and UV monitoring revealed a $\sim$64\,d period, which has been considered as the binary orbital period \citep[][here after MPS+14]{MotchPS2014}. The optical peak was interpreted as the illumination of the companion star by the X-ray emission. The light curve modeling constrained the black hole mass as $M_{\rm{BH}}=3.45-15M_{\odot}$ and an orbital eccentricity of $e=0.27-0.41$. Further analyzing the phase jitter of the optical maximum implied that \ngc\ may exhibit a superorbital modulation with a period of $1,800-3,200$\,d (MPS+14). 

Recently, P13 was found to host an accreting pulsar with a spin period of $\sim$0.42\,s and a period derivative of $\dot{P}\sim-3.5\times10^{-11}$\,s\,s$^{-1}$ \citep{FuerstWH2016, IsraelPE2016}. This discovery implies that the Roche lobe of the supergiant is much larger than the value reported in MPS+14, and the Roche-lobe filling accretion can only occur when the neutron star passes the periastron with an eccentricity of $e=0.46-0.55$ \citep{IsraelPE2016}. If it is true, we would see the orbital X-ray modulation since the mass accretion rate is orbital-phase dependent. Moreover, no significant period derivative due to the orbital Doppler effect was found in the long \emph{NuSTAR} observation \citep{FuerstWH2016}. Hence, another possible origin of the 64\,d optical period is the superorbital modulation instead of orbital modulation like the 55--62\,d quasi-periodic modulation in M82~X-2 \citep{PashamS2013, QiuGW2015, KongHL2016} and the 78-d period in NGC 5907 ULX1 \citep{WaltonFB2016}. 

Here we report the detection of the 65\,d X-ray periodicity in P13 with a \swift\ monitoring. The data selections of XRT and UVOT are stated in Section \ref{observation}. We describe the timing analysis, including the detection of this periodic signal, the significant test, and further phase evolution analysis in Section \ref{analysis}. Then we discuss possible orbital/superorbital solutions and their implications in Section \ref{discussion} and summarize this research in Section \ref{summary}.

\begin{figure*}
\plotone{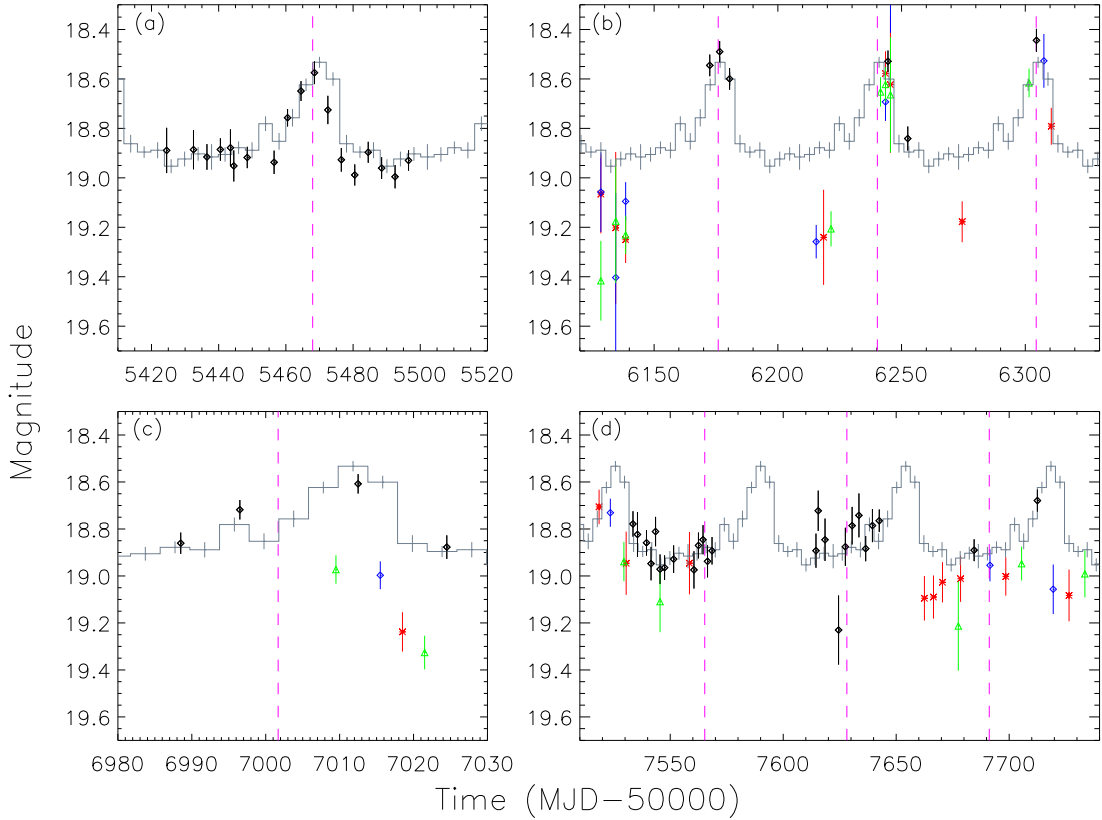}
\caption{One-day binned UVOT light curves of P13 in \emph{u} (black), \emph{w1} (blue), \emph{m2} (red), and \emph{w2} (green) bands. This source was monitored by UVOT in (a) 2010, (b) mid-2011 to early-2013, corresponding to X-ray quiescence, (c) late-2014, and (d) from 2016 April to December. Magenta vertical dashed lines indicate the expected arrival time for the optical peak by MPS+14, while the gray histogram is the \emph{u}-band light curve folded with 64.24\,d. \label{lc_uvot}}
\end{figure*}

\section{\emph{SWIFT} Observations \label{observation}}
\setcounter{footnote}{0}
P13 has been monitored with \swift\ since 2010 August. Four series of regular monitoring have been made in 2010, mid-2011 to early-2013, late-2014, and from 2016 April until December. We extracted the 0.3--10\,keV XRT light curve from all the photon counting mode observations and binned it per snapshot via the online XRT product generator\footnote{\url{http://www.swift.ac.uk/user\_objects/}} \citep{EvansBP2007,EvansBP2009}. It first localized the source position using the first observation if the source is observable. Then, it defined a background annulus around the source and excluded other sources lying in the background region. Only those data points with higher than 3$\sigma$ detection significance were used in the following analysis. Therefore, data points between mid-2011 to early-2013 were excluded.  We further applied a 3-$\sigma$ criterion to reject those data with extremely high uncertainties. A total of 221 snapshots were collected with a time span of 2,309\,d. Figure~\ref{lc_xrt}(a) shows the XRT light curves, while the light curves for individual epochs are shown in Figure~\ref{lc_xrt}(b)--(d).

For the UVOT data, we downloaded all the 55 \emph{u}, 15 \emph{uw1}, 20 \emph{um2}, and 17 \emph{uw2} bands observations from the \texttt{HEASARC} data archive. The \emph{Swift}-specific task \texttt{uvotmaghist} in the \texttt{HEAsoft} (version 6.19) was used to extract UVOT light curves with the \swift\ UVOT \texttt{CALDB} (version 20160321). A source aperture radius of 3\arcsec\ as recommended by the \swift\ manual\footnote{\url{http://swift.gsfc.nasa.gov/analysis/threads/uvot_thread_aperture.html}} and a 15\arcsec\ radius circular source-free background were adopted. We noted that the local field around the target is crowded and full of diffuse emission of the host galaxy, so the background region was chosen far away from the target to minimize contamination. This likely overestimates the target's magnitude (i.e., star + diffuse emissions). It is acceptable because we focus on timing analysis and an absolute photometry is not necessary. Figure~\ref{lc_uvot} shows the UVOT light curves at different epochs. 

\section{Timing Analysis \label{analysis}}
\subsection{X-ray Periodicity}
\begin{figure}
\plotone{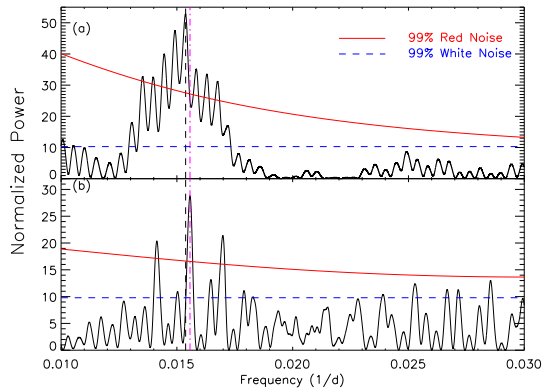}
\caption{Zoom-in view of the Lomb-Scargle power spectra of (a) \swift\ XRT and (b) UVOT \emph{u}-band light curves of P13. Blue-dashed and red lines are the 99\% significance levels of white- and red-noises, respectively. The vertical black dashed line is the best-determined X-ray period of $P=65.05$\,d, and the magenta dash-dotted line is the best-determined \emph{u}-band period of $P=64.24$\,d. \label{lsp}}
\end{figure}

We noticed that the X-ray light curve of P13 has a long-term trend that may contaminate the power spectrum (Figure~\ref{lc_xrt}).  Hence, we removed the linear trend by subtracting the light curve with the averaged count rate in each segment (Figure~\ref{lc_xrt}b--d). Then, we applied the Lomb-Scargle periodogram \citep{Lomb1976, Scargle1982} to search for periodicity. The best-determined period is $P_X=65.05\pm0.10$\,d\BD{, while the uncertainty is estimated using the formula in \citet{Horne1986}}. \BD{A series of alias caused by the gap and several harmonics can be seen. The zoom-in view of the power spectrum is shown in Figure~\ref{lsp}a in order to compare with the power spectrum of the optical data.} We also used the epoch folding algorithm to search for periods and estimated the uncertainty using the $\chi^2$ test \citep{Leahy1987}. The \BD{period and corresponding uncertainty are} consistent with those obtained from the Lomb-Scargle method. We further estimated the 99\% white noise significance level by adopting the empirical function in \citet{Horne1986}, and the 99\% red noise level by fitting the time series with the {\tt REDFIT} algorithm \citep{SchulzM2002, FarrellBS2009}. The signal is well above the red-noise significance levels.

\begin{figure}
\plotone{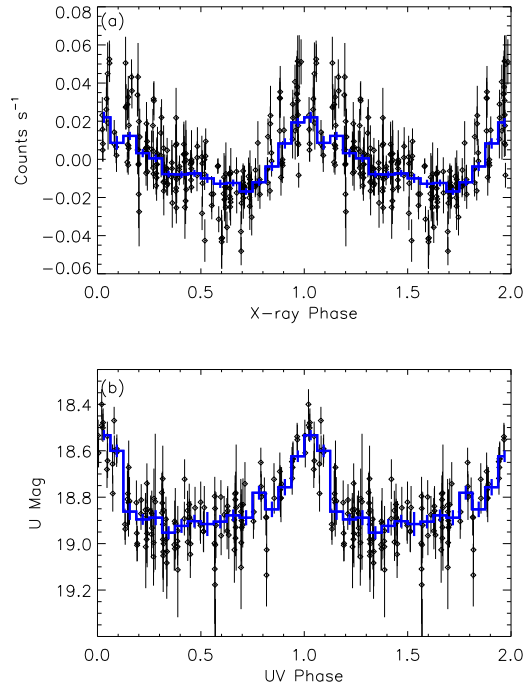}
\caption{Folded (a) XRT and (b) \emph{u}-band light curves of P13 with both data points of each snapshot (black diamonds) and binned profile (blue histogram). The folding periods are $65.05$\,d and $64.24$\,d for the X-ray band and the \emph{u}-band, respectively.\BD{ We set the phase 0 as the peaks of X-ray and optical bands individually.}  \label{fold_lc}}
\end{figure}

The X-ray folded light curve is shown in Figure~\ref{fold_lc}a. The profile is asymmetric, showing a fast rise ($\gtrsim$0.2\,cycles) and a slow decay ($\gtrsim$0.4\,cycles). We overlaid the folded light curve on the original XRT light curves (Figure~\ref{lc_xrt}b--d) and shifted the count rate according to the trend. This matches the data well although the amplitude of the data in panel (b) seems to be lower than that in panel (d). This probably indicates a connection between the modulation amplitude and the X-ray flux level. 

\subsection{Optical Periodicity}
We then performed the same analysis on the \emph{u}-band optical light curve (Figure~\ref{lc_uvot}). The resulting Lomb-Scargle power spectrum is shown in Figure~\ref{lsp}(b). The best-determined period is $P_{\rm{opt}}=64.24\pm0.13$\,d. The power is lower than that in X-rays, but still well above the noise levels. \BD{We also tried the epoch folding analysis and yielded $P_{\rm{opt}}=64.32\pm0.10$\,d.} We noted that the optical period differs from the X-ray one at $\gtrsim 3\sigma$ level. However, the difference becomes less significant if we consider the most conservative uncertainty described by the Fourier width, i.e., $\sigma_P=P^2/2T$, where $T$ is the length of the time baseline. This yields $\sim$0.9\,d uncertainties for both the X-ray and optical periods. Moreover, the phase of the optical peak may shift due to the superorbital modulation (MPS+14). Therefore, detailed phase evolution analysis is necessary to further investigate their relationship. The \emph{u}-band folded light curve is shown in Figure~\ref{fold_lc}(b). The optical peak can be well fit by a Gaussian function with $\sigma$ of $\sim$4.8\,d.

\begin{figure*}
\plottwo{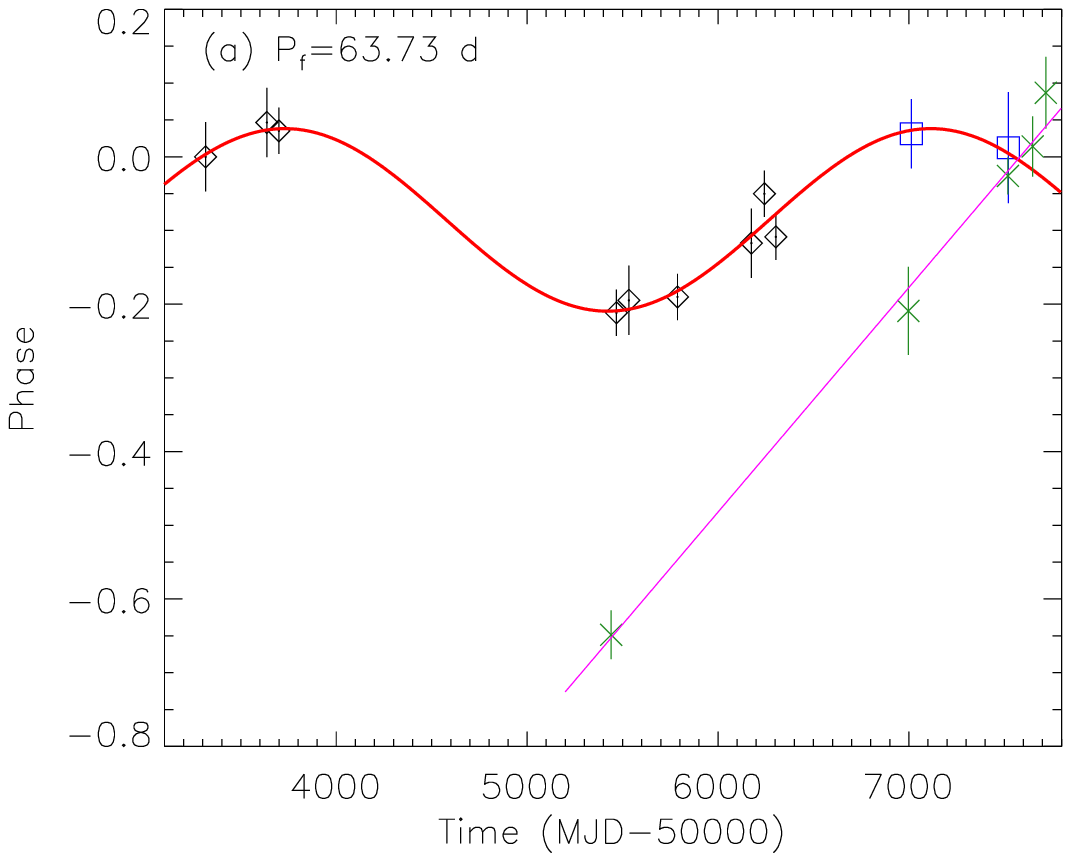}{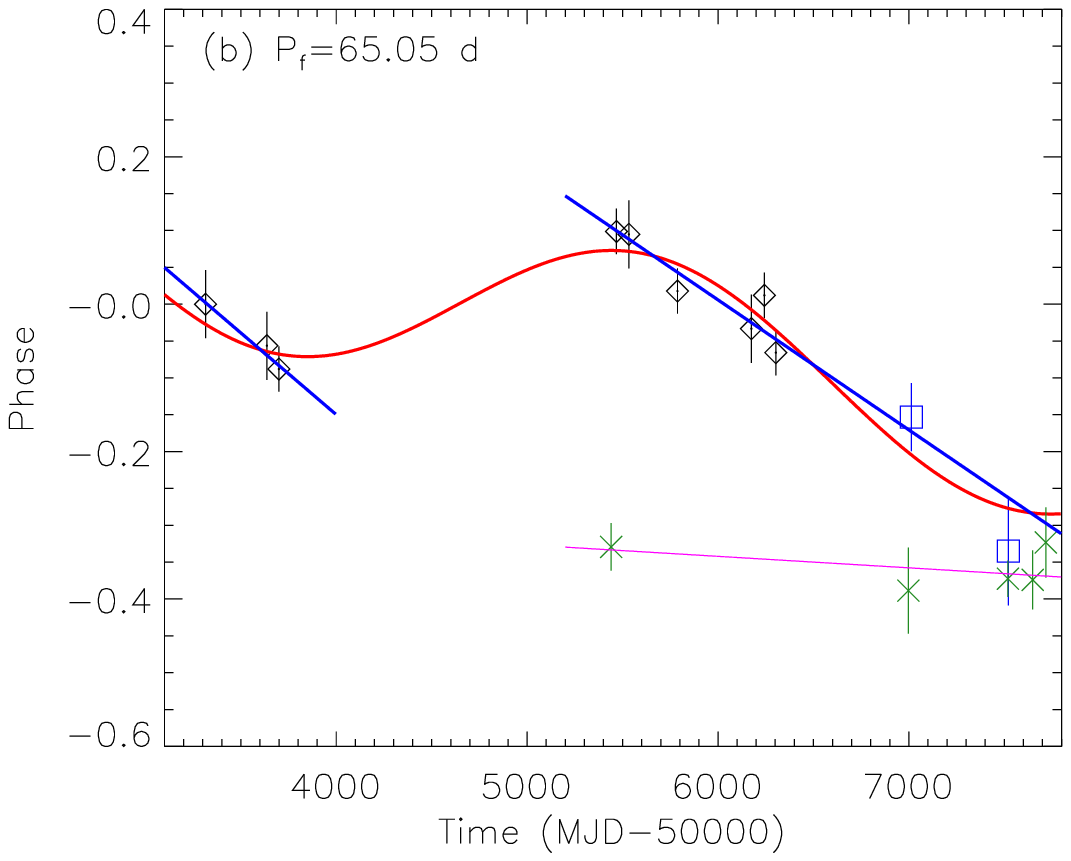}
\caption{Evolution of the optical and X-ray peak arrival phases of P13 with a folding period of (a) $P_f=63.73$\,d, and (b) $P_f=65.05$\,d. Black diamonds are optical peak arrival phases adopted from MPS+14, blue squares are optical data obtained with \swift\ UVOT data in 2015--2016, and green crosses are X-ray peak arrival phases obtained with \swift\ XRT. Red curves are best-fit superorbital modulations with periods of (a) 3400\,d and (b) 3900\,d\BD{, while the blue straight lines in panel (b) are the best linear fits before 2006 and after 2010.} Magenta lines are the best linear fit of X-ray peak arrival times with a period of $P=64.99$\,d. \label{o_c}}
\end{figure*}

\subsection{Phase Evolution \label{phase_evo}}
To further distinguish the X-ray and optical periods, we analyzed their phase evolution. We adopted the times of photometric maximum in the \emph{V}-band observed by the Las Campanas and ESO VLT, \emph{u}-band and multi-band estimation by \swift\ UVOT from MPS+14. After 2013, the cadence of the \emph{u}-band light curve is not enough to clearly identify the optical maxima. Following the multi-band identification in MPS+14, we found an optical maximum around MJD 57,013 according to the change of the flux. Moreover, we observed a decreasing wing before $\sim$MJD 57,545. Assuming that the optical profile is stable, we estimated that there is an optical maximum at $\sim$MJD 57,522. We assigned a $4.8$\,d uncertainty corresponding to the Gaussian width of the optical peak as a conservative estimate. 

Comparing to the optical peak arrival times, the X-ray ones are more difficult to estimate due to the asymmetric profile. We therefore used the folded light curve as a template, and fit the light curve \BD{of individual cycles} by using the Levenberg-Marquardt non-linear fitting algorithm {\tt　MPFIT}\footnote{\url{http://purl.com/net/mpfit}} \citep{Markwardt2009}. During the fitting, we allowed \BD{three} parameters to vary: the time of X-ray maximum, which is defined as phase 0 in the folded light curve, the modulation amplitude, and a constant term. We determined four arrival times of the X-ray peak as: MJD$55,440\pm2$\,d, $56,998\pm4$\,d, $57,519\pm2$\,d, $57,649\pm3$\,d, and $57,717\pm3$\,d. 

\BD{To investigate the evolution of the measured optical and X-ray periods in detail, we tried two approaches. We set the phase zero at MJD 55,314.8, corresponding to the first arrival time of the optical maximum in MPS+14 to compare both the X-ray and optical phase evolution in the same standard.}

\BD{Based on $P_{\rm{opt}}=63.52$\,d (MPS+14), we refined this solution with additional two UVOT data points by fitting the phase evolution of the optical peak arrival phases with a sinusoidal function. The best-fit result indicates that $P_{\rm{opt}}=63.73\pm0.1$\,d and $P_{\rm{sup}}=3400\pm400$\,d, with $\chi_\nu^2=0.6$ for 7 degrees of freedom (dof), consistent with the 68\% confidence interval reported by MPS+14. We then used the folding period$P_f=P_{\rm{opt}}=63.73$\,d to plot the X-ray peak arrival phases, and checked if the X-ray and optical periods have the same evolutionary pattern (Figure~\ref{o_c}a). We found that the X-ray data show a significant phase drift, which can be well described by a straight line with a constant period of $P_X=64.99\pm0.07$\,d. In this case, $P_{\rm{opt}}$ can be interpreted as the beat period of $P_X$ and $P_{\rm{sup}}$. The expected superorbital modulation period is $P_{\rm{sup}}=1/(1/P_{\rm{opt}}-1/P_X)=3200\pm350$\,d, consistent with that obtained from the phase jitter.}

\BD{We further calculated the phase evolution of the X-ray maxima based on $P_f=P_X=65.05$\,d, the result determined using the Lomb-Scargle periodogram from the \swift\ XRT data. The X-ray data were well fit by a linear function with a period identical to previous case. Then, we plotted the optical peak arrival phases according to this linear X-ray ephemeris to obtain the evolution of the optical data (see Figure~\ref{o_c}b). We found that the optical data can be well fit with a sinusoidal curve indicating that $P_{\rm{opt}}=64.82\pm0.1$\,d and $P_{\rm{sup}}=3900\pm800$\,d with a worse but still acceptable $\chi_\nu^2=1.2$. This is not very far from the alternating acceptable solution in MPS+14. Moreover, the optical period of this solution is roughly consistent with the X-ray one although the X-ray data did not show the sinusoidal phase jitter. We also tried to divide the optical data into two epochs, before 2006 and after 2010, and fit them with a straight line individually. The best fit results are $P_{\rm{opt}}\textrm{(before~2006)}=64.1\pm0.6$\,d and $P_{\rm{opt}}\textrm{(after~2010)}=64.3\pm0.1$\,d. The fit after 2010 is acceptable with a $\chi_\nu^2=0.8$ and 6 dof. These two periods are consistent with the optical period derived from the Lomb-Scargle periodogram of the \swift\ \emph{u}-band light curve. } 


\section{Discussion \label{discussion}}

Our timing analysis of \swift\ XRT data reveals a stable $\sim$65\,d X-ray modulation in \ngc. This period is possibly different from the optical period. The optical period was attributed to the orbital period (MPS+14) although this is not conclusive \citep{FuerstWH2016}. If instead, the X-ray modulation is orbital in origin, an orbital-phase-dependent accretion rate could naturally explain the data. \BD{On the other hand, the $\sim$65\,d period could be the superorbital period similar to the $\sim$55--62\,d modulation in M82~X-2 \citep{PashamS2013, QiuGW2015, KongHL2016} and the $\sim$78\,d period in NGC 5907 ULX1 \citep{WaltonFB2016}. We here discuss their implications according to these two possible origins.}

\BD{If $P_X$ is the orbital period, $P_{\rm{opt}}$ can be interpreted as the beat period of $P_X$ and $P_{\rm{sup}}$ for the case in Figure~\ref{o_c}a. Similar behavior was observed in Galactic X-ray binary systems, e.g., the dipping low-mass X-ray binary 4U 1916$-$053, in which the optical modulation period is $\sim$1\% longer than its orbital period \citep{ChouGB2001, HuCC2008}.  In the case of \ngc, the disk has to have retrograde precession like Her X-1 \citep{GerendB1976, OgilvieD2001} because $P_{\rm{opt}}<P_X$. If this is the case, the enhancement of the optical emission could be dominated by the illumination of the tilted accretion disk. 

Another scenario is that both the $P_{\rm{opt}}$ and $P_X$ represent the orbital modulation but the optical one shows phase jitter with a much longer period (Figure~\ref{o_c}b). The scenario proposed by MPS+14 interpreted the phase jitter as the effect of the disk precession like Her X-1. In this model, the optical variability is caused by the shadow of the tilted accretion disk. However, the drift of the optical maximum is likely linear instead of sinusoidal in Her X-1 \citep{DeeterCG1976, GerendB1976}, in which the optical maximum drifted from orbital phase $\sim$0.8 to $\sim$0.2 in one superorbital cycle. Taking P13 as an analogous case of Her X-1, the superorbital period could be estimated as $\sim$2700\,d by calculating the time interval between two zero-crossings of the drifting trend (blue lines in Figure~\ref{o_c}b). If it is true, the orbital period will dominate the power spectrum of the optical data as long as the time span is much longer than one superorbital cycle. It could possibly explain the strongest $65.15$\,d peak in the power spectrum presented in MPS+14.}


\BD{Both the aforementioned cases require a superorbital modulation to explain the phase drift or the period difference. A currently acceptable explanation of the superorbital modulation is the precession of a tilted and warped accretion disk (MPS+14). The stability of a radiation-driven warping of an accretion disk} could be described by two parameters: the binary separation $r_b$ in units of $GM_1/c^2$ and the mass ratio $q$, where $M_1$ is the mass of the compact object \citep{OgilvieD2001}. In P13, the binary separation has to be larger than $R_2\approx$50--60\,$R_{\odot}$ for a B9Ia supergiant with a mass of 18--23\,$M_{\odot}$ \citep{IsraelPE2016}, implying that $r_b/10^6>17$. The mass ratio is expected to be $q>10$ since the compact object is a neutron star. Accordingly, P13 lies on the upper-right corner of the $r_b-q$ plot \citep[see Figure~7 of][]{OgilvieD2001}, beyond the stable zone for steady precession. Therefore, the disk precession period of \ngc\ is probably aperiodic like Cyg X-2 or quasi-periodic like SMC X-1 \citep{ClarksonCC2003b,ClarksonCC2003a}. The (instantaneous) precession period can vary between neighboring cycles \citep{Trowbridge2007, Hu2011}. If one of $P_{\rm{opt}}$ and $P_X$ is the orbital period and another one is caused by beating, the beat period is expected to be less stable. Hence, further \swift\ multi-band monitoring could help us to test the stability and explore the evolution of these two periodicities. 

\BD{If both $P_X$ and $P_{\rm{opt}}$ are superorbital periods, the orbital period should be much shorter. The relationship between the orbital and superorbital modulation periods of Galactic HMXBs are summarized in \citet{CorbetK2013}. Both the Roche-lobe filled (disk-fed) and the wind-fed HMXBs show a positive correlation between $P_{\rm{orb}}$ and $P_{\rm{sup}}$, but the superorbital modulation period in disk-fed systems ($P_{\rm{sup}}/P_{\rm{orb}}\sim$10--20) are systematically longer than those in the wind-fed systems ($P_{\rm{sup}}/P_{\rm{orb}}\sim$2--4). If the relationship also hold for the ultraluminous HMXBs, P13 could have an oribtal period of 3--7\,d for disk-fed assumption, or 16--32\,d for wind-fed accretion scenario. In addition, if we plot the spin period against the orbital period of known HMXBs, wind-fed and disk-fed systems show clear different distributions \citep{Corbet1986, BildstenCC1997, LiSL2016}. The spin periods of disk-fed systems as well as M82~X-2 anti-correlate with the orbital periods. If \ngc\ is a disk-fed system, the orbital period estimated above agrees this relationship. On the other hand, \citep{KarinoM2016} suggested that M82~X-2 is possibly a short-period extension of the standard Be-type HMXBs. If both measured $P_X$ and $P_{\rm{opt}}$ are superorbital periods, \ngc\ could be in a similar regime since the spin and superorbital modulation periods are similar to those in M82~X-2.}

\BD{The phase drift or the discrepancy between $P_X$ and $P_{\rm{orb}}$ are intriguing if the $\sim$65\,d period is due to disk precession. It is probably the effect of a third companion or other unknown mechanisms.} Further monitoring of the spin period with \emph{XMM-Newton} or \emph{NuSTAR} could unambigously reveal the true orbital period if the orbital Doppler effect can be observed. As a result, a complete picture of this system can be well established.

\section{Summary \label{summary}}
With the \swift\ XRT monitoring of \ngc, we determined an X-ray period of $\sim$65\,d. The X-ray profile is asymmetric and has a much wider peak than the optical profile. Through the phase evolution, we proposed two possible \BD{combinations} of the optical and X-ray periods. \BD{If both the X-ray and optical periods represent the orbital period, the phase drift may be interpreted as the effect of a superorbital modualtion originated from the precession of the disk. The optical modulation period could also be the beat period of the orbital and the disk precession periods. In both cases, the disk precession period seems to be unstable according to the radiation-driven warping model. Another possibility is that the $\sim$65\,d modulation is the superorbital modulation. An orbital period of 3--7\,d could be inferred according to the $P_{\rm{spin}}$--$P_{\rm{orb}}$ and $P_{\rm{orb}}$--$P_{\rm{sup}}$ relations if P13 is a disk-fed system. In this case, the much longer modulation with a period of thousands of days could be originated from other unknown mechanisms. Therefore, further monitoring of \ngc\ is required to probe the origin of these two periods. }

\acknowledgments
\BD{We would like to thank the referee for the valuable comments and suggestions.} This work made use of data supplied by the UK Swift Science Data Centre at the University of Leicester. CPH and CYN are supported by a GRF grant of Hong Kong Government under HKU 17300215P. AKHK is supported by the Ministry of Science and Technology of the Republic of China (Taiwan) through grants 103-2628-M-007-003-MY3 and 105-2112-M-007-033-MY2.

\facilities{\emph{Swift}}

\bibliography{../reference}

\begin{thebibliography}{}
\expandafter\ifx\csname natexlab\endcsname\relax\def\natexlab#1{#1}\fi
\providecommand{\url}[1]{\href{#1}{#1}}

\bibitem[{{Bachetti} {et~al.}(2014){Bachetti}, {Harrison}, {Walton},
  {Grefenstette}, {Chakrabarty}, {F{\"u}rst}, {Barret}, {Beloborodov}, {Boggs},
  {Christensen}, {Craig}, {Fabian}, {Hailey}, {Hornschemeier}, {Kaspi},
  {Kulkarni}, {Maccarone}, {Miller}, {Rana}, {Stern}, {Tendulkar}, {Tomsick},
  {Webb}, \& {Zhang}}]{BachettiHW2014}
{Bachetti}, M., {Harrison}, F.~A., {Walton}, D.~J., {et~al.} 2014, \nat, 514,
  202

\bibitem[{{Bildsten} {et~al.}(1997){Bildsten}, {Chakrabarty}, {Chiu}, {Finger},
  {Koh}, {Nelson}, {Prince}, {Rubin}, {Scott}, {Stollberg}, {Vaughan},
  {Wilson}, \& {Wilson}}]{BildstenCC1997}
{Bildsten}, L., {Chakrabarty}, D., {Chiu}, J., {et~al.} 1997, \apjs, 113, 367

\bibitem[{{Chou} {et~al.}(2001){Chou}, {Grindlay}, \& {Bloser}}]{ChouGB2001}
{Chou}, Y., {Grindlay}, J.~E., \& {Bloser}, P.~F. 2001, \apj, 549, 1135

\bibitem[{{Clarkson} {et~al.}(2003{\natexlab{a}}){Clarkson}, {Charles}, {Coe},
  \& {Laycock}}]{ClarksonCC2003b}
{Clarkson}, W.~I., {Charles}, P.~A., {Coe}, M.~J., \& {Laycock}, S.
  2003{\natexlab{a}}, \mnras, 343, 1213

\bibitem[{{Clarkson} {et~al.}(2003{\natexlab{b}}){Clarkson}, {Charles}, {Coe},
  {Laycock}, {Tout}, \& {Wilson}}]{ClarksonCC2003a}
{Clarkson}, W.~I., {Charles}, P.~A., {Coe}, M.~J., {et~al.} 2003{\natexlab{b}},
  \mnras, 339, 447

\bibitem[{{Corbet}(1986)}]{Corbet1986}
{Corbet}, R.~H.~D. 1986, \mnras, 220, 1047

\bibitem[{{Corbet} \& {Krimm}(2013)}]{CorbetK2013}
{Corbet}, R.~H.~D., \& {Krimm}, H.~A. 2013, \apj, 778, 45

\bibitem[{{Deeter} {et~al.}(1976){Deeter}, {Crosa}, {Gerend}, \&
  {Boynton}}]{DeeterCG1976}
{Deeter}, J., {Crosa}, L., {Gerend}, D., \& {Boynton}, P.~E. 1976, \apj, 206,
  861

\bibitem[{{Evans} {et~al.}(2007){Evans}, {Beardmore}, {Page}, {Tyler},
  {Osborne}, {Goad}, {O'Brien}, {Vetere}, {Racusin}, {Morris}, {Burrows},
  {Capalbi}, {Perri}, {Gehrels}, \& {Romano}}]{EvansBP2007}
{Evans}, P.~A., {Beardmore}, A.~P., {Page}, K.~L., {et~al.} 2007, \aap, 469,
  379

\bibitem[{{Evans} {et~al.}(2009){Evans}, {Beardmore}, {Page}, {Osborne},
  {O'Brien}, {Willingale}, {Starling}, {Burrows}, {Godet}, {Vetere}, {Racusin},
  {Goad}, {Wiersema}, {Angelini}, {Capalbi}, {Chincarini}, {Gehrels}, {Kennea},
  {Margutti}, {Morris}, {Mountford}, {Pagani}, {Perri}, {Romano}, \&
  {Tanvir}}]{EvansBP2009}
---. 2009, \mnras, 397, 1177

\bibitem[{{Farrell} {et~al.}(2009){Farrell}, {Barret}, \&
  {Skinner}}]{FarrellBS2009}
{Farrell}, S.~A., {Barret}, D., \& {Skinner}, G.~K. 2009, \mnras, 393, 139

\bibitem[{{Feng} \& {Soria}(2011)}]{FengS2011}
{Feng}, H., \& {Soria}, R. 2011, \nar, 55, 166

\bibitem[{{F{\"u}rst} {et~al.}(2016){F{\"u}rst}, {Walton}, {Harrison}, {Stern},
  {Barret}, {Brightman}, {Fabian}, {Grefenstette}, {Madsen}, {Middleton},
  {Miller}, {Pottschmidt}, {Ptak}, {Rana}, \& {Webb}}]{FuerstWH2016}
{F{\"u}rst}, F., {Walton}, D.~J., {Harrison}, F.~A., {et~al.} 2016, \apjl, 831,
  L14

\bibitem[{{Gerend} \& {Boynton}(1976)}]{GerendB1976}
{Gerend}, D., \& {Boynton}, P.~E. 1976, \apj, 209, 562

\bibitem[{{Horne} \& {Baliunas}(1986)}]{Horne1986}
{Horne}, J.~H., \& {Baliunas}, S.~L. 1986, \apj, 302, 757

\bibitem[{{Hu} {et~al.}(2008){Hu}, {Chou}, \& {Chung}}]{HuCC2008}
{Hu}, C.-P., {Chou}, Y., \& {Chung}, Y.-Y. 2008, \apj, 680, 1405

\bibitem[{{Hu} {et~al.}(2011){Hu}, {Chou}, {Wu}, {Yang}, \& {Su}}]{Hu2011}
{Hu}, C.-P., {Chou}, Y., {Wu}, M.-C., {Yang}, T.-C., \& {Su}, Y.-H. 2011, \apj,
  740, 67

\bibitem[{{Israel} {et~al.}(2016){Israel}, {Belfiore}, {Stella}, {Esposito},
  {Casella}, {De Luca}, {Marelli}, {Papitto}, {Perri}, {Puccetti}, {Rodriguez
  Castillo}, {Salvetti}, {Tiengo}, {Zampieri}, {D'Agostino}, {Greiner},
  {Haberl}, {Novara}, {Salvaterra}, {Turolla}, {Watson}, {Wilms}, \&
  {Wolter}}]{IsraelBS2016}
{Israel}, G.~L., {Belfiore}, A., {Stella}, L., {et~al.} 2016, arXiv:1609.07375

\bibitem[{{Israel} {et~al.}(2017){Israel}, {Papitto}, {Esposito}, {Stella},
  {Zampieri}, {Belfiore}, {Rodr{\'{\i}}guez Castillo}, {De Luca}, {Tiengo},
  {Haberl}, {Greiner}, {Salvaterra}, {Sandrelli}, \& {Lisini}}]{IsraelPE2016}
{Israel}, G.~L., {Papitto}, A., {Esposito}, P., {et~al.} 2017, \mnras, 466, L48

\bibitem[{{Karachentsev} {et~al.}(2003){Karachentsev}, {Grebel}, {Sharina},
  {Dolphin}, {Geisler}, {Guhathakurta}, {Hodge}, {Karachentseva}, {Sarajedini},
  \& {Seitzer}}]{KarachentsevGS2003}
{Karachentsev}, I.~D., {Grebel}, E.~K., {Sharina}, M.~E., {et~al.} 2003, \aap,
  404, 93

\bibitem[{{Karino} \& {Miller}(2016)}]{KarinoM2016}
{Karino}, S., \& {Miller}, J.~C. 2016, \mnras, 462, 3476

\bibitem[{{Kong} {et~al.}(2016){Kong}, {Hu}, {Lin}, {Li}, {Jin}, {Liu}, \&
  {Yen}}]{KongHL2016}
{Kong}, A.~K.~H., {Hu}, C.-P., {Lin}, L.~C.-C., {et~al.} 2016, \mnras, 461,
  4395

\bibitem[{{Leahy}(1987)}]{Leahy1987}
{Leahy}, D.~A. 1987, \aap, 180, 275

\bibitem[{{Li} {et~al.}(2016){Li}, {Shao}, \& {Li}}]{LiSL2016}
{Li}, T., {Shao}, Y., \& {Li}, X.-D. 2016, \apj, 824, 143

\bibitem[{{Lomb}(1976)}]{Lomb1976}
{Lomb}, N.~R. 1976, \apss, 39, 447

\bibitem[{{Markwardt}(2009)}]{Markwardt2009}
{Markwardt}, C.~B. 2009, in ASP Conf. Ser. 411, Astronomical Data Analysis
  Software and Systems XVIII, ed. D.~A. {Bohlender}, D.~{Durand}, \&
  P.~{Dowler}, (San Francisco, CA: ASP), 251

\bibitem[{{Motch} {et~al.}(2011){Motch}, {Pakull}, {Gris{\'e}}, \&
  {Soria}}]{MotchPG2011}
{Motch}, C., {Pakull}, M.~W., {Gris{\'e}}, F., \& {Soria}, R. 2011, AN, 332,
  367

\bibitem[{{Motch} {et~al.}(2014){Motch}, {Pakull}, {Soria}, {Gris{\'e}}, \&
  {Pietrzy{\'n}ski}}]{MotchPS2014}
{Motch}, C., {Pakull}, M.~W., {Soria}, R., {Gris{\'e}}, F., \&
  {Pietrzy{\'n}ski}, G. 2014, \nat, 514, 198 (MPS14)

\bibitem[{{Ogilvie} \& {Dubus}(2001)}]{OgilvieD2001}
{Ogilvie}, G.~I., \& {Dubus}, G. 2001, \mnras, 320, 485

\bibitem[{{Pannuti} {et~al.}(2011){Pannuti}, {Schlegel}, {Filipovi{\'c}},
  {Payne}, {Petre}, {Harrus}, {Staggs}, \& {Lacey}}]{Pannuti2011}
{Pannuti}, T.~G., {Schlegel}, E.~M., {Filipovi{\'c}}, M.~D., {et~al.} 2011,
  \aj, 142, 20

\bibitem[{{Pasham} \& {Strohmayer}(2013)}]{PashamS2013}
{Pasham}, D.~R., \& {Strohmayer}, T.~E. 2013, \apjl, 774, L16

\bibitem[{{Poutanen} {et~al.}(2007){Poutanen}, {Lipunova}, {Fabrika},
  {Butkevich}, \& {Abolmasov}}]{PoutanenLF2007}
{Poutanen}, J., {Lipunova}, G., {Fabrika}, S., {Butkevich}, A.~G., \&
  {Abolmasov}, P. 2007, \mnras, 377, 1187

\bibitem[{{Qiu} {et~al.}(2015){Qiu}, {Liu}, {Guo}, \& {Wang}}]{QiuGW2015}
{Qiu}, Y., {Liu}, J., {Guo}, J., \& {Wang}, J. 2015, \apjl, 809, L28

\bibitem[{{Radburn-Smith} {et~al.}(2011){Radburn-Smith}, {de Jong}, {Seth},
  {Bailin}, {Bell}, {Brown}, {Bullock}, {Courteau}, {Dalcanton}, {Ferguson},
  {Goudfrooij}, {Holfeltz}, {Holwerda}, {Purcell}, {Sick}, {Streich}, {Vlajic},
  \& {Zucker}}]{Radburn-SmithJS2011}
{Radburn-Smith}, D.~J., {de Jong}, R.~S., {Seth}, A.~C., {et~al.} 2011, \apjs,
  195, 18

\bibitem[{{Read} \& {Pietsch}(1999)}]{ReadP1999}
{Read}, A.~M., \& {Pietsch}, W. 1999, \aap, 341, 8

\bibitem[{{Scargle}(1982)}]{Scargle1982}
{Scargle}, J.~D. 1982, \apj, 263, 835

\bibitem[{Schulz \& Mudelsee(2002)}]{SchulzM2002}
Schulz, M., \& Mudelsee, M. 2002, Comput. Geosci., 28, 421

\bibitem[{{Shakura} \& {Sunyaev}(1973)}]{ShakuraS1973}
{Shakura}, N.~I., \& {Sunyaev}, R.~A. 1973, \aap, 24, 337

\bibitem[{{Trowbridge} {et~al.}(2007){Trowbridge}, {Nowak}, \&
  {Wilms}}]{Trowbridge2007}
{Trowbridge}, S., {Nowak}, M.~A., \& {Wilms}, J. 2007, \apj, 670, 624

\bibitem[{{Tully} {et~al.}(2016){Tully}, {Courtois}, \& {Sorce}}]{TullyCS2016}
{Tully}, R.~B., {Courtois}, H.~M., \& {Sorce}, J.~G. 2016, \aj, 152, 50

\bibitem[{{Walton} {et~al.}(2016){Walton}, {F{\"u}rst}, {Bachetti}, {Barret},
  {Brightman}, {Fabian}, {Gehrels}, {Harrison}, {Heida}, {Middleton}, {Rana},
  {Roberts}, {Stern}, {Tao}, \& {Webb}}]{WaltonFB2016}
{Walton}, D.~J., {F{\"u}rst}, F., {Bachetti}, M., {et~al.} 2016, \apjl, 827,
  L13

\end{thebibliography}

\end{document}